\documentclass[12pt,titlepage]{utarticle}
\usepackage{amsmath,amssymb,dcpic,pictexwd,graphicx,color}
\usepackage[rflt]{floatflt}
\usepackage{hyperref}

\numberwithin{equation}{section}

\def\defeq{\buildrel\rm def\over=}

\def\Et{\widehat{E}}
\def\Ht{\widehat{H}}
\def\blg{\Omega E_8}
\def\flg{LE_8}
\def\flG{LG}

\def\clg{\widetilde{LE}^k_8}
\def\ulg{\widetilde{LE}^{k=1}_8}

\def\hlg{\widehat{LE_8}}
\def\Tlg{LSpin(11)}
\def\hlG{\widehat{LG}}

\def\Lg{L\mathfrak{g}}
\def\cLg{\widetilde{L\mathfrak{g}}}

\def\wdg{{\wedge}}                              

\newcommand{\pxpy}[2]{\frac{\partial{#1}}{\partial{#2}}}

\def\BZ{\mathbb{Z}}

\def\BR{\mathbb{R}}
\def\BT{U(1)}
\def\MB{\mathbf{B}}
\def\ME{{\mathbf{E}}}
\def\Sp{{\bf S}^{\bf 1}_p}
\newcommand{\MS}[1]{{{\bf S}^{#1}}}               
\renewcommand{\CP}[1]{{\mathbb{C}{\mathbf{P}^{#1}}}}        
\renewcommand{\arraystretch}{1.25}

\def\mcO{\mathcal{O}}
\def\mcA{\mathcal{A}}

\def\Tr{\mathrm{Tr}}

\def\Ot{\widehat{\Omega}}

\def\ie{\textit{i.e.,\ }}

\begin{document}
\preprint{UTTG--06--04\\
\texttt{hep-th/0406218}\\}

\title{Loop Groups, Kaluza-Klein Reduction and M-Theory}

\author{Aaron Bergman and Uday Varadarajan}

\oneaddress{Theory Group, Physics Department\\
             University of Texas at Austin\\
             Austin, TX 78712\\ {~}\\
             \email{abergman@physics.utexas.edu}
             \email{udayv@physics.utexas.edu}}

\Abstract{ We show that the data of a principal $G$ bundle over a principal
  circle bundle is equivalent to that of a $\hlG \defeq \BT \ltimes
  \flG$ bundle over the base of the circle bundle. We apply this to
  the Kaluza-Klein reduction of M-theory to IIA and show that certain
  generalized characteristic classes of the loop group bundle encode
  the Bianchi identities of the antisymmetric tensor fields of IIA
  supergravity. We further show that the low dimensional
  characteristic classes of the central extension of the loop group
  encode the Bianchi identities of massive IIA, thereby adding support
  to the conjectures of hep-th/0203218.  }

\maketitle
\newpage

\section{Introduction and Summary}\label{sec:intro}

It has long been established that IIA supergravity is the dimensional
reduction of the unique theory of supergravity in eleven dimensions.
We have learned in the last decade, however, that eleven dimensional
supergravity is the low energy limit of a still mysterious theory
termed M-theory and that the Kaluza-Klein reduction of this theory is
the ten dimensional type IIA superstring. Furthermore, it has been
understood that the charges and fluxes associated with antisymmetric
tensor fields in these supergravity theories obey certain quantization
conditions which can differ from the usual Dirac quantization
condition \cite{Witten:1997md, Moore:1999gb}. Often, it turns out that
one can find geometric objects that model these `quantized' fluxes.
The relevant object for M-theory turns out to be somewhat complicated
\cite{Diaconescu:2003bm}, but an essential ingredient is an
$E_8$-bundle that models an element in $H^4(M,\BZ)$.\footnote{In fact,
  any model for $K(\BZ,4)$ will do in the construction of
  \cite{Diaconescu:2003bm}. $E_8$ bundles are satisfyingly
  geometrical, however, and they will be the focus of this paper. It
  is curious to note that one such alternate model of $K(\BZ,4)$ in
  low dimensions is $\ulg$ bundles, where $\ulg$ is the universal
  central extension of the loop group of $E_8$. Of course, $\ulg$
  appears in the world sheet description of the heterotic string,
  which is dual to M-theory on manifolds with boundary
  \cite{Horava:1996qa,Horava:1996ma}. }  This $E_8$-bundle was first
introduced in \cite{Witten:1997md} and was motivated by the appearance
of an $E_8$ gauge symmetry on boundaries in M-theory as described in
\cite{Horava:1996qa,Horava:1996ma}.

One can now ask, what is the relation of the quantization condition of
the M-theory 4-form to that of the RR and NS forms in type IIA?
The local relationship between the classical p-form fields and
gauge invariances in M-theory and those of type IIA supergravity have
long been understood as that of Kaluza-Klein reduction. However, the
relationship between their quantization conditions is a subtle issue.  For
example, the Bianchi identity of type IIA supergravity, $dG_4 + G_2
\wedge H = 0$, already shows that in the presence of nonzero $H$ the
field $G_4$ is not a closed form.  Thus, unlike the 4-form in
M-theory, integral cohomology does not seem to be an appropriate
context for understanding its quantization condition.  
Indeed, the description of the quantization condition of
RR-fluxes in the type IIA superstring at vanishingly small $g_s$ has
been related to (twisted) K-theory \cite{Moore:1999gb} rather than
ordinary integral cohomology.

Given the duality between M-theory compactified on a circle of radius
$R_{11} \propto g_s$ and IIA string theory with coupling $g_s$, we
can rephrase the above question differently. From the
perspective of type IIA string theory, we wish to understand the
finite $g_s$ behavior of the quantization condition of type
IIA fluxes. On the other hand, from the M-theory perpective we wish to
understand how the $E_8$-bundle data associated with the quantization
of the M-theory 4-form relates to the quantization condition of its
Kaluza-Klein descendents when we take the M-theory circle to be very
small.

The compatibility of K-theory in IIA and the $E_8$-bundle picture in
M-theory was first examined in \cite{Diaconescu:2000wy} by comparing
topological flux contributions to the partition functions of M-theory
and IIA on compact manifolds related by dimensional reduction. These
comparisons were made for terms associated with vanishing $H$ flux
(equivalently, for M-theory four forms which could be pulled back from
IIA) and for $G_0=0$, as no framework exists for incorporating
nonzero $G_0$ into M-theory data. They established that certain
subtle topological phases in these partition functions agreed and that
the quantization conditions associated with $E_8$ gauge bundles in
M-theory were, in this sense, compatible with the K-theoretic
quantization of fluxes IIA. However, their calculation indicated that
there did {\em not} seem to exist a 1-1 correspondence between
quantized flux configurations in M-theory and IIA. Rather, it was only
after performing a {\em sum} over configurations in the partition
functions that agreement was observed. It was therefore suggested that the
relationship between the antisymmetric tensor fields in M-theory and
IIA must be understood as a {\em quantum} equivalence.

Here, we take a very modest step towards better understanding this
relationship by describing how loop groups of $E_8$ can allow one to
organize and generalize the Kaluza-Klein reduction of the quantization
data in M-theory to include all nontrivial RR and NS fluxes, even $G_0
\neq 0$.  To do this, we will consider in detail the Kaluza-Klein
reduction of the $E_8$-bundle data and certain generalizations
suggested by this process. It was conjectured \cite{HoravaUnp} that
the resulting data might be understood in terms of bundles of a loop
group of $E_8$ and that such a description might be related to the
K-theory description in a more transparent way, perhaps along the
lines suggested in \cite{Bouwknegt:2000qt}. This idea was first
explored in \cite{Adams:2002rx} and later in
\cite{Mathai:2003mu,Evslin:2003kn}, where it was suggested that for
trivial M-theory circle bundles, the dimensional reduction of the
$E_8$-bundle in M-theory would be a $\flg$-bundle over the ten
dimensional base (related ideas were further developed in
\cite{Evslin:2003hd, Evslin:2003wb, Bouwknegt:2003vb,
Kriz:2004xj, JAU}). Here, $\flg$
is the free loop group of $E_8$ defined as maps from $\MS{1}$ into
$E_8$ with pointwise multiplication. However, we will find that this
conjecture must be modified in order to include the case of a
nontrivial M-theory principal circle bundle. We will establish a
correspondence between $E_8$-bundles in M-theory and $\hlg$-bundles in
IIA, where $\hlg$ is a slightly modified version of the loop group
given by $\hlg \defeq \BT \ltimes \flg$. 

It was further conjectured in \cite{Adams:2002rx} that massive IIA
supergravity \cite{Romans:1986tz} would be related to the centrally
extended loop group $\clg$. If this could be verified and an
appropriate loop group generalization of the $\eta$ invariants used in
the work of \cite{Diaconescu:2000wy, Moore:2002cp} defined, one might
be able to extend their results to $G_0 \neq 0$ by adding to the
M-theory partition function sectors corresponding to the $\clg$
quantization of massive IIA for all $k = G_0$.\footnote{For an
alternative viewpoint on incorporating massive IIA backgrounds in
M-theory see \cite{Hull:1998vy, Haack:2001iz, Moore:2002cp}.} While
such a task is far beyond the scope of this work, we will show that
the characteristic classes of $\clg$ bundles can reproduce the Bianchi
identities of massive IIA (thereby explaining many of the results of
\cite{JAU}).  A further simple consequence \cite{JAU} of the $\clg$
quantization is that D6-brane charge is valued in $\BZ_k$ in massive
IIA. We describe how this result can be understood via a
St\"{u}ckelberg mechanism and is verified in the work of
\cite{Polchinski:1996sm} on Calabi-Yau compactifications of massive
IIA. Finally, using this correspondence, we will describe a simple
example of the puzzles that arise when one attempts to compare the the
K-theory classification of fluxes directly with the loop group bundle
picture.

The outline of our paper is as follows. In section \ref{sec:bundle},
we will show directly how to construct this $\hlg$ bundle as a special
case of a more general construction which relates $G$ bundles on a
principal circle bundle $Y$ to $\hlG$ bundles on its base $X$. In
section \ref{sec:charclass}, we will discuss the characteristic
classes for this bundle. We will show that the topological data of the
$\hlg$ bundle in low dimensions is faithfully encoded in a pair of
classes, one in ordinary cohomology and the other in a generalized
cohomology theory inspired by the Gysin sequence. In particular, we
will see that these classes necessarily obey the Bianchi identities of
IIA supergravity. We will also show that a similar relationship holds
between the characteristic classes of $\clg$ bundles and the Bianchi
identities of massive IIA supergravity.  In order to establish these
results, it will be necessary both to compute the cohomology of the
classifying space of $\hlg$ bundles and to explore the generalized
cohomology theory alluded to earlier. This will be relegated to
section \ref{sec:cohom} and the appendices to spare the reader who
wishes to avoid such details.  Finally, we consider some interesting
examples and comment on the relation between the loop group bundles
explored in this paper and the K-theory classification given in
\cite{Moore:1999gb} which was fruitfully exploited in
\cite{Diaconescu:2000wy, Moore:2002cp}. This and other issues will be
the subject of section \ref{sec:disc}.

Previous work relating to the appearance of Kac-Moody symmetries
in Kaluza-Klein reduction has appeared in \cite{Dolan:1984aa,Dolan:1984fm}.

\section{$\hlG$ Bundles and Kaluza-Klein Reduction}\label{sec:bundle}


The basic setup we will consider is diagrammed in Figure
\ref{basicsetup}. We have an $(n+1)$-manifold $Y$ which is a $\BT$
\textit{principal} bundle\footnote{\label{foot1} More generally, we
could have a $\MS{1}$ fiber bundle with structure group
Diff($\MS{1}$). Much of the following construction is
identical. However, in the end, instead of a principal $\hlG$-bundle
we would obtain a $\MS{1} \times \flG$ bundle with structure group
$\mathrm{Diff}(\MS{1}) \ltimes \flG$. We do not know how to
characterize such bundles.  One cannot help but notice the relation of
this sort of symmetry to that in CFT. We do not know this significance
of this.}  over a simply connected $n$-manifold $X$, and a $G$-bundle
$E$ over $Y$. In the specific case of the reduction of M-theory to
type IIA string theory, $Y$ is eleven dimensional, $X$ is ten
dimensional, and we consider $G=E_8$. This can be summarized in the
following diagram:\newline
\begin{center}
\begin{minipage}{2.5in}
\begin{equation}
\label{thebundle}
\begin{split}
\begindc{\commdiag}
         \obj(3,1)[X]{$X$}
         \obj(1,3)[T]{$\BT$}
         \obj(3,3)[Y]{$Y$}
         \obj(1,5)[E8]{$G$}
         \obj(3,5)[E]{$E$}
         \mor{T}{Y}{}
         \mor{Y}{X}{$\rho$}
         \mor{E}{Y}{$\pi$}
         \mor{E8}{E}{}
\enddc
\end{split}\ .
\end{equation}
\end{minipage}
\hspace{1in}
\begin{minipage}{2.5in}
\includegraphics[height=2in,width=2in]{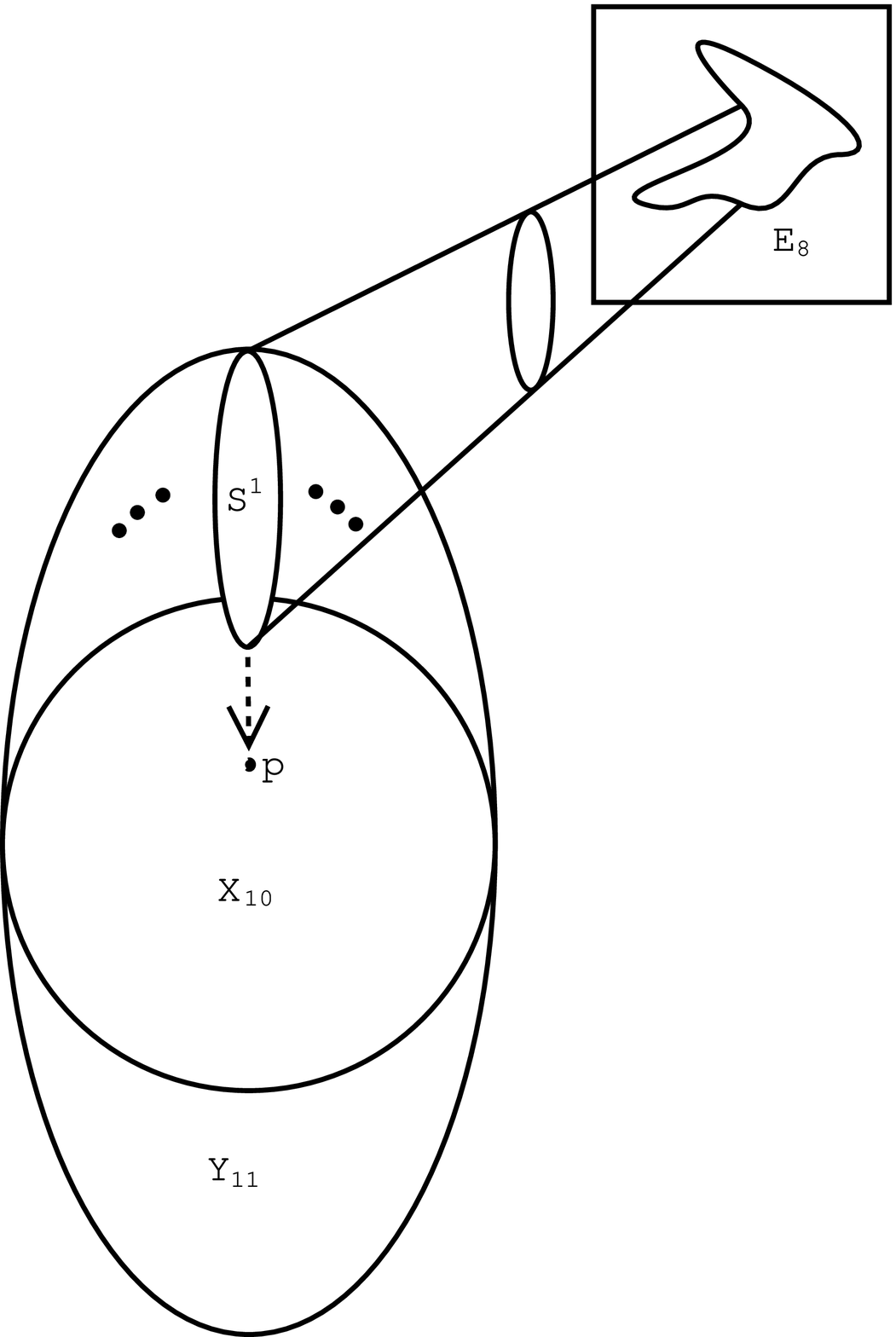}
\\[5mm]
\stepcounter{figure}
Figure \thefigure: The basic setup
\label{basicsetup}
\end{minipage}
\end{center}

Before giving the correct construction of the $\hlG$ bundle, let us
see where the attempt to construct a $\flG$ bundle goes wrong. Over
any point $p \in X$, $\rho^{-1}(p)$ is a circle, $\Sp \subset Y$. As
long as G is connected, any $G$-bundle over a circle is trivial and
has global sections. Thus, for each $p$ we consider the space of
sections, $\Gamma(\rho^{-1}(p), E) = \Gamma(\Sp, E)$.  The triviality
of the bundle over $\Sp$ allows us to think of this as maps from $\Sp$
to $G$. Taking the union of these spaces for all $p \in X$, one might
think that we have formed a principal $\flG$ bundle, as the free loop
group is given by $\mathrm{Maps}(\MS{1} \rightarrow G)$. However, a
principal $\flG$ bundle also possesses a free left $\flG$ action. As
the multiplication on $\flG$ is pointwise, in order to define a global
$\flG$ action, we would need to identify the points of the $\Sp$ for
all $p \in X$ with the points of the single circle in the group
$\flG$. This immediately implies that there exists global sections of
the original $\MS{1}$ bundle. This is only possible when $Y$ is a
trivial bundle, \ie $Y \cong X \times \MS{1}$.  In fact, if we choose
any global section of $Y$, we can use the $\BT$ action on $Y$ to give
an identification of each $\Sp$ with the original the group $\BT$
where the section defines an identity. So, in essence, what we lack is
a choice of an identity on each $\Sp$.\footnote{More precisely, the
problem is that $\Sp$ is a $\BT$ \textit{torsor} rather than the group
$\BT$ itself.  The choice of an identity identifies the torsor with a
copy of the group.}

Lacking this identity in general, we will simply adjoin one to the
construction.\footnote{We are indebted to Dan Freed for suggesting
this.} In other words, over each point $p \in X$, we take the space
$\Sp \times \Gamma(\Sp, E)$ where the first factor corresponds to all
possible choices of identity. This is topologically $\MS{1} \times
\flG$.  Taking the union for all points $p$ gives a bundle which we
will denote $\Et$.

We now show that this is a principal bundle. Given the point $s \in
\Sp$, the $\BT$ action on $Y$ gives us an identification of $\Sp$ with
$\BT$. We use this to let $\flG$ act on $\Gamma(\Sp,E)$. Finally,
there is a $\BT$ action on the $\MS{1}$ part of the fiber of $\Et$
which acts exactly as does the $\BT$ action on the fiber of $Y$. This
group action is, in fact, a semidirect product, $\hlG = \BT \ltimes
\flG$. To see this we note that for each point $s \in \Sp$, the $\BT$
action on $Y$ gives us a particular identification of $\Sp$ with $\BT$
such that $s$ is mapped to the identity element in $\BT$. Using this
map, we can associate to any $t \in \Sp$ an angle $\theta$ by the
relation $t=s+\theta \in \Sp$, where we denote the action of $\BT$ on
$\Sp$ by addition. Thus, we can rewrite $(s,f(t)) \in \Sp \times
\Gamma(\Sp, E)$ as $(s,f(s + \theta))$. This allows us to define the
action of $h_1 = (\phi_1,g_1) \in \hlG$ on $\Et$ by,
\begin{equation}
(s,f(t)) \cdot h_1 = g_1 (s,f(s+\theta)) \cdot (\phi_1,g_1) \defeq
  (s+\phi_1,f(s+\theta) \cdot g_1(\theta)).
\end{equation}
To exhibit the group law in $\hlG$, we act again with
$(\phi_2, g_2) \in \hlG$. In particular, if we define $s'$ and
$\theta'$ by the relations,
\begin{equation}
s' = s+\phi_1  ~~~~~~~ \theta' = \theta - \phi_1,
\end{equation}
we find that
\begin{align}
(s+\phi_1,f(s+\theta) \cdot g_1(\theta)) \cdot (\phi_2,g_2) & = 
(s',f(s'+\theta') \cdot g_1(\theta)) \cdot (\phi_2,g_2) \\
 & = (s'+\phi_2,f(s'+\theta') \cdot g_1(\theta) \cdot g_2(\theta')).
\end{align}
From this we deduce that the group law for $\hlG$ must be,
\begin{equation}
(\phi_1, g_1) \cdot (\phi_2,g_2) = (\phi_1+\phi_2, g_1(\theta)
  \cdot g_2 (\theta')) = (\phi_1+\phi_2, g_1(\theta) \cdot g_2 (\theta
  - \phi_1)),
\end{equation}
which defines the semidirect product $\BT \ltimes \flG$ where the
action of $\BT$ on $\flG$ is given by rotation of the loop.

We can summarize this construction as the following correspondence:
\begin{equation}
\label{bundlemap}
\begin{split}
\begin{minipage}{3in}
\begindc{\commdiag}
         \obj(3,1)[X]{$X$}
         \obj(1,3)[T]{$\BT$}
         \obj(3,3)[Y]{$Y$}
         \obj(1,5)[G]{$G$}
         \obj(3,5)[E]{$E$}
         \mor{T}{Y}{}
         \mor{Y}{X}{}
         \mor{E}{Y}{}
         \mor{G}{E}{}
\enddc
\end{minipage}
\qquad\Longleftrightarrow\qquad
\begin{minipage}{3in}
\begindc{\commdiag}
         \obj(3,1)[X]{$X$}
         \obj(1,3)[LG]{$\hlG$}
         \obj(3,3)[Et]{$\Et$}
         \mor{LG}{Et}{}
         \mor{Et}{X}{}
\enddc
\end{minipage}
\end{split}
\end{equation}
In fact, as we will show in section \ref{sec:cohom}, this
correspondence is invertible: given an $\hlG$ bundle, we can
uniquely construct an $E_8$ bundle on the total space of its associated
principal circle bundle.\newline

\section{The $\hlg$ Bundle and Its Characteristic
  Classes}\label{sec:charclass}

As mentioned in the introduction, the particular case of $G = E_8$ is
particularly relevant for M-theory.  In light of the results of the
previous section, it is interesting to ask if there might be a
similarly fruitful relationship between $\hlg$ bundles on $X_{10}$ and
the antisymmetric tensor fields of type IIA. That such a relationship
should exist is certainly not transparent from the string theory point
of view as the quantization conditions for the RR forms of IIA are
only known to be easily expressed in terms of classes in K-theory.

In this section, we will only begin to explore this question by asking
if such a relationship can pass the most rudimentary test of
consistency: do the characteristic classes of $\hlg$ bundles regarded
as elements of de Rham cohomology obey the type IIA supergravity
Bianchi identities? We will see that this is indeed the case. As
further evidence for the relevance of loop groups of $E_8$ to IIA, we
will also demonstrate that the characteristic classes of the centrally
extended loop group are consistent with the Bianchi identities of
massive IIA.

\subsection{Supergravity and Characteristic Classes}

Characteristic classes are certain elements in the cohomology of a
manifold that characterize, often incompletely, principal bundles over
that manifold.  For example, for a $U(N)$-bundle, these are the Chern
classes. Or, for a $\BT$-bundle, the sole characteristic class is the
Euler class\footnote{This is just the first Chern class in a flimsy
  disguise.}, $e$. As we will review in the next section, an
$E_8$-bundle on $Y_{11}$ is {\em completely} classified in low dimensions by
a class in the fourth integer cohomology of the base. This is an
analogous to the well known result that $U(1)$-bundles are classified
by the Euler class.  Making use of both these facts, the topological
data in (\ref{thebundle}) is completely encoded in the following
information:
\begin{equation} \label{charclasses}
e \in H^2(X_{10},\BZ) \quad \mathrm{and} \quad
a \in H^4(Y_{11},\BZ).
\end{equation}
Physically, these classes should correspond to topological fluxes in
IIA and M-theory assoicated with $G_2$, the RR 2-form field strength
in IIA string theory and $G$, the M-theory 4-form. However, as
described in \cite{Witten:1997md}, the correspondence is not quite
what one might naively expect from (\ref{charclasses}). While it is
true that $e = [G_2]$ in $H^2_\mathrm{DR}(X_{10})$, the correct
quantization condition for the 4-form is $a=[G] - \frac{\lambda}{2}
\in H^4(Y_{11},\BZ)$ where $\lambda = p_1(Y_{11})/ 2$. This issue is
properly dealt with in \cite{Diaconescu:2003bm}. For the physical
interpretation of what follows, we will assume that $p_1(Y_{11}) = 0$.
Nonetheless, as there is a map to $\MB E_8$ in
\cite{Diaconescu:2003bm}, we believe that some variation of our
construction will be relevant in the general case.\footnote{In our
  picture, it is reasonable to conjecture that these gravitational
  corrections can be properly treated by adding the $Spin(11)$ frame
  bundle of $Y_{11}$ to the $E_8$ bundle and considering the
  dimensional reduction of the data of the full $G=E_8 \times
  Spin(11)$ bundle. By the results of the previous section, this would
  suggest a description in IIA in terms of $\hlG = \BT \ltimes (\flg
  \times \Tlg)$ bundles, where the $\BT$ acts by simultaneous rotation
  of the loop in both factors.}

In order to compare to IIA, however, we would like express the
data of the $E_8$-bundle on $Y_{11}$ in terms of geometric data on
$X_{10}$.\footnote{As $a$ classifies $E_8$-bundles, at the level of
  cohomology this problem is solved by a long exact sequence known as
  the Gysin sequence, as we will review in section \ref{sec:cohom}.
  The loop group picture, however, provides a geometric interpretation
  for the result and generalizes to massive IIA.}  As the results of
the previous section imply that the data in (\ref{charclasses})
classify $\hlg$ bundles over $X_{10}$ as well, one might guess that
this information can be expressed in terms of the characteristic classes of the
$\hlg$ bundle.  To find the characteristic classes of the $\hlg$
bundle, we begin by asking if the data in (\ref{charclasses}) provide
natural candidates for them.  Of course, as $e=[G_2]$ is already
expressed in terms of data on $X_{10}$, we can just reinterpret it as
the characteristic class of the $\hlg$ bundle associated with the
$\BT$ of $\hlg$. As for $a$, the only natural cohomology class on
$X_{10}$ that one can obtain from it is via the pushforward map which
relates $H^4(Y_{11})$ to $H^3(X_{10})$.  More explicitly, in terms of
the M-theory four form $G$ we have $H = \rho_*(G)$ where the
pushforward map is given by integration over the $\MS{1}$ fiber, and
$H$ is interpreted as the NS 3-form field strength in IIA.  In fact,
there is an interesting interpretation of $H$ (which we will return to
later) as the obstruction to lifting the $\hlg$ bundle to a $\BT
\ltimes \ulg$ bundle.

As we will see in the next section, $[H]$ and $[G_2]$ are the only two
characteristic classes of $\hlg$ bundles in ordinary cohomology in low
dimensions.  However, it turns out that these two classes are not
independent as $H$ is constrained by the fact that it arises as a
pushforward of $G$. The nature of this constraint is best illustrated
by the following simple example.  Consider $E_8$-bundles over
$Y=\MS{3} \times \MS{3}$. Since $H^4(Y)=0$ they must all be
trivial. Now, consider the dimensional reduction of $Y$ along the
circle of a Hopf fibration of the first $\MS{3}$ factor to $X = \MS{2}
\times \MS{3}$. Clearly, there are nontrivial elements $[H] \in
H^3(X)$ which do not lift to classes $[G] \in H^4(Y)$ in $Y$.  Indeed,
any $G$ flux which has a nonzero pushforward, $H = \rho_*(G)$, would
have to be nontrivial through a cycle of the form $\MS{3} \times
\MS{1}$ where $\MS{1}$ is the fiber. However, the Hopf fibration over
$\MS{2}$ renders the cycle homologically trivial in $Y$. Now, as $G_2$
detects the Hopf fibration of the circle over the $S^2$, this
situation is characterized by the fact that $\int_X H \wedge G_2 =
\int_{\MS{3}} H \int_{\MS{2}} G_2 \neq 0$. Generalizing this example,
if there exists a 5-cycle like $X$ in $X_{10}$ such that the $H$ flux
restricted to $X$ is non-trivial through a 3-cycle, and that the $G_2$
flux restricted to $X$ is non-trivial through an intersecting 2-cycle
so that $\int_X H \wedge G_2 \neq 0$, a similar argument would show
that $H$ does not lift to a $G$ flux in $Y_{11}$. In fact, as we will
show in section \ref{sec:cohom}, the class $[H]$ lifts to some $[G]
\in Y_{11}$ if and only if $[H \wedge G_2] = 0$. Thus, the low
dimensional characteristic classes of $\hlg$ obey $[H \wedge G_2]=0$.

This presents us with a puzzle, however, if we were to hope that all
the information about the $E_8$-bundle on $Y_{11}$ would be contained
in these characteristic classes. The pushforward $\rho_*$ has a kernel
including forms pulled back from the base. One can certainly have
elements of $H^4(Y_{11})$ in that kernel. For example, consider an
$E_8$-bundle over $Y=\MS{1} \times \MS{4}$ which has $n$ units of
instanton number on the $\MS{4}$. Then, $[G]$ is just $n$ times the unit
volume form on $\MS{4}$. If we consider the dimensional reduction to
$X = \MS{4}$, $H=\int_{\MS{1}} G = 0$ and so $[G]$ is clearly in the
kernal of the pushforward map. Thus, the characteristic classes we
have found will not detect that part of the classification of $E_8$
bundles. In other words, there are distinct $E_8$-bundles that give
rise to identical $[H]$ and $[G_2]$, so these two characteristic
classes cannot alone classify $\hlg$ bundles.

An obvious place to look for more information is the RR
4-form defined by
\begin{equation}
\label{g4def}
G_4 = \rho_*(\mcA \wdg G),
\end{equation}
where $G$ is again the M-theory four form and we have introduced
$\mcA$, an Ehresmann connection\footnote{This is also the global
  angular form of \cite{MR83i:57016}.} on $Y_{11}$.
While $G_4$ is gauge invariant, it is not closed and instead obeys a
modified Bianchi identity:
\begin{equation}
\label{g4ident}
dG_4 + G_2\wdg H = 0\ .
\end{equation}
While the metric in M-theory does give us an explicit choice of a
connection\footnote{In particular, if the metric on $Y_{11}$ is $ds^2
  = ds^2_\mathrm{base} + (d\theta + \alpha)^2$, then $\mcA \propto
  d\theta + \alpha$.}, the $\hlg$ bundle is topological, and its
characteristic classes should not depend on any such choice. We would
like to find a mathematical context for understanding how $G_4$ might
classify $\hlg$ bundles. 

As we discuss in section \ref{sec:cohom}, to do this we need to
explore in greater detail the relationship between the cohomology ring
of $Y_{11}$ and that of $X_{10}$. The cohomology groups of $Y_{11}$
and $X_{10}$ fit into a cohomology long exact sequence known as the
Gysin sequence.\footnote{The Gysin sequence was used in a similar
context in \cite{Mathai:2003mu}.}
In section \ref{sec:cohom} we will show how this
sequence can be used to describe $a \in H^4(Y_{11}, \BZ)$ in terms of a
class in a generalized cohomology theory on $X_{10}$ which captures
the information in both $G_4$ and $H$. While we will postpone a
detailed treatment of this issue to that section, we will describe
here the relevant results for a de Rham version of the theory.

We begin with a complex of differential forms\footnote{
We first encountered this complex in \cite{MR2002j:53028} which
refers to \cite{MR86i:57033}. All the proofs in this paper are due 
to us.}
\begin{equation}
\label{doubform}
\Ot^i(X_{10}) \defeq \Omega^i(X_{10}) \oplus \Omega^{i-1}(X_{10})\ .
\end{equation}
Furthermore, we will choose an explicit representative of the Euler
class, $e$, of the M-theory circle bundle. As above, this is just the
RR 2-form. Given such a choice, $G_2$, we define the differential:
\begin{equation}
d(\alpha,\beta) = (d\alpha + G_2\wdg\beta,-d\beta)
\end{equation}
for $(\alpha,\beta)\in\Ot^i(X_{10})$. This squares to zero, so we can
pass to the cohomology which we will denote $\Ht^i_{e}(
X_{10})$. It is not hard to see that the complexes for different
choices of the representative of the Euler class are quasiisomorphic,
\ie have the same cohomology, justifying the notation. We will prove
in the next section that the cohomology of this complex is, in fact,
$H^4(Y_{10},\BR)$.

$(G_4,H)$ is closed in this complex, as long as we have\footnote{Note
  that these differ from the usual equations for a connection by a
  factor of $2\pi$. We do this to save some space. The second
  condition is always possible because $d\rho_*(\mcA) =
  -\rho_*\rho^*(G_2) = 0$, so $\rho_*(\mcA)$ is a constant function.
  Thus, the condition just reflects a choice of normalization.}
\begin{equation}
\label{connecteqn}
d\mcA = -\rho^*G_2\  \quad \mathrm{and} \quad \rho_*(\mcA) = 1\ .
\end{equation}
Thus, it defines an element in the cohomology. One can verify that the
class so obtained only depends on the cohomology class of $G$. We claim
that this is the appropriate context for the IIA 4-form.

Given a cohomology class $[(G_4,H)]$, we can reconstruct the
original class in $H^4_\mathrm{DR}(Y_{11})$ as follows. First, we must 
choose a connection. Then, we define the form:
\begin{equation}
\label{gdef}
G = \rho^*(G_4) - \mcA\wdg\rho^*H\ .
\end{equation}
$G$ is closed, so it defines a class in $H^4_{\mathrm{DR}}(Y_{11})$. It is
an easy exercise to see that it does not depend on the addition of an
exact `form' to $(G_4,H)$. Finally, let us show that it does not
depend on the choice of $\mcA$. We choose two different connection
$\mcA_1$ and $\mcA_2$. Then,
\begin{equation}
G^1 - G^2 = (\mcA_2 - \mcA_1)\wdg\rho^*H\ .
\end{equation}
From (\ref{connecteqn}), we have
\begin{equation}
d(\mcA_2 - \mcA_1) = -\rho^*G_2 + \rho^*G_2 = 0\ .
\end{equation}
Now, because we have chosen a nontrivial circle bundle and a simply
connected base, $X_{10}$, we have $H^1(Y_{11},\BR) = 0$. This can be
seen, for example, from the Gysin sequence for the fibration.  Since
the cohomology is trivial, the difference in connections must be
exact, \ie $\mcA_2 - \mcA_1 = dB$, giving $G^1 - G^2 =
d(B\wdg\rho^*H)$. This means that the cohomology class of $G$ is
independent of our choice of connection, and we are done.

Thus, we see that $e = [G_2] \in H^2(X_{10})$ and $[(G_4,H)] \in
\Ht^4_{e}(X_{10})$ are characteristic classes for the $\hlg$ bundle
and contain all the information in the setup (\ref{thebundle}). The
Bianchi identities for the RR two and four forms in type IIA
supergravity follow from these definitions. This provides a
mathematical context for these well-known equations.

\subsection{The Romans Mass}\label{sec:romans}

Now, we turn to the situation where $G_0 = k$ is nonzero. This is the
situation in massive IIA supergravity. Here, the Bianchi identity for
$G_2$ is modified so that it obeys
\begin{equation}
\label{romang2}
dG_2 = G_0 H\,
\end{equation}
in close analogy with the identity $dG_4 + G_2 \wedge H = 0$ we
encountered above. Since $G_2$ is no longer a closed form, it cannot
be the Euler class of a circle bundle, and the meaning of M-theory
becomes somewhat unclear. It was conjectured in \cite{Adams:2002rx}
that in this situation we should look at $\clg$ bundles over $X_{10}$
where $\clg$ is the level $k$ central extension of $\flg$.
Topologically, the group $\flg \cong E_8 \times \blg$, so
$H^2(\flg,\BZ) = \BZ$. Thus, there exist circle bundles over $\flg$
labeled by the integers. The group $\clg$ is topologically the circle
bundle over $\flg$ with Euler class $k$ times the generator of
$H^2(\flg,\BZ)$. 

Using this fact, we can construct the first characteristic class of
$\clg$ bundles. The homotopy exact sequence for the above circle
fibration, shows us that $\pi_1(\clg) = \BZ_k$. Thus, we can use a
simple obstruction theory argument\footnote{See, for example,
  \cite{Witten:1986bt}.}  to give us a characteristic class $h \in
H^2(X_{10},\BZ_k)$. Note that since $\pi_1(\ulg) = 0$, in analogy with
the second Steifel Whitney class, this class can be interpreted as the
obstruction to lifting a given $\clg$ bundle to a bundle of its
universal covering group $\ulg$.\footnote{In fact \cite{MR88i:22049},
  we can construct $\clg$ from $\ulg$ by taking the quotient $\clg
  =\frac{ \ulg \times \BT } { \{ (\theta, -k \theta) \in \BT_K \times
    \BT \} }$ where $\BT_K$ is the central element in $\ulg$.} 
In order to motivate the relation of $h$ to supergravity fields, first
consider some cochain model $C^*(X_{10}, \BZ)$ for integral
cohomology.  The most obvious way to get an element $h \in
H^2(X_{10},\BZ_k)$ in such a model is to take the mod $k$ reduction
of a closed cochain $\chi_2 \in C^2(X_{10},\BZ)$
which represents a class in $H^2(X_{10},\BZ)$. However, since we are
working mod $k$, we really only need that $\chi_2$ is closed mod
$k$. That is, we only require that $\chi_2$ obeys
\begin{equation} \label{romang3}
d \chi_2 = k \chi_3,
\end{equation}
for some cocycle $\chi_3$. Note that (\ref{romang3}) defines a natural map
$\beta: H^2(X_{10}, \BZ_k) \rightarrow H^3(X_{10},\BZ)$ by
$\beta(h)=[\chi_3]$ which is called the Bockstein homomorphism. 
We will associate $\beta(h)=[H]$ with the quantized $H$-flux.
Since $k \beta(h)= k [\chi_3]  = [d \chi_2] = 0$, we see that
$\beta(h)=[H]$ must correspond to a $\BZ_k$ torsional class in
$H^3(X,\BZ)$. In particular, this means that we can find a de Rham
representative, $kH$, of $k\beta(h)$. Of course, de Rham cohomology
does not detect torsion, so $H$, as a form, is exact even when not
multiplied by $k$.\footnote{Interestingly, torsion three forms have a
  simple geometric model in terms of $SU(N)/\BZ_N$ bundles or Azumaya
  algebras \cite{Witten:1998cd,Kapustin:1999di}.}
Further, note that the kernal of $\beta(h)$ consists precisely of
mod $k$ reductions of elements of $H^2(X_{10},\BZ)$. Clearly, we
would like to associate this information with $G_2$. However, just as
in the previous section, this is complicated by the modified Bianchi
identity (\ref{romang2}).

To find a model for $G_2$, we use the analogy between (\ref{romang3})
and the modified Bianchi identity of $G_4$ in the previous section to
construct a modified cochain complex for computing
$H^i(X_{10},\BZ_K)$. The cochains consist of pairs $(\alpha,\beta)$
with $\alpha \in C^{i}(X_{10},\BZ)$ and $\beta \in
C^{i+1}(X_{10},\BZ)$, and the differential is defined by,
\begin{equation}
d(\alpha, \beta) = (d \alpha - k \beta, d \beta).
\end{equation} 
In analogy with the results described in the previous section, the
cohomology of this complex computes $H^i(X_{10},\BZ_k)$. Now,
comparing with (\ref{romang2}) shows that we can interpret the pair
$(G_2,H)$ as a 2-cocycle in a de Rham version of the above complex.
Note that when $H=0$, $G_2$ is closed and would naively represent an
arbitrary element of integral cohomology. The quantization condition
coming from $\clg$ suggests a nontrivial modification of this naive
expectation, however. In fact, it is only the mod $k$ reduction of this
element of integral cohomology that is physically relevant.
As we will discuss in section \ref{sec:disc}, this expectation is
borne out by certain examples discussed in
\cite{Polchinski:1996sm}.

Massive IIA supergravity also has a four form which obeys the following
Bianchi identity:
\begin{equation}
d G_4 = G_2 \wedge H = \frac{1}{G_0} G_2 \wedge dG_2\ .
\end{equation}
This implies
\begin{equation}
\widetilde{G_4} =  G_4 - {\frac{1}{2G_0}} G_2 \wedge G_2
\end{equation}
is closed. We believe that $\widetilde{G_4}$ should be thought of as a 
characteristic class for the $\clg$ bundle. The existence of such
a class is proven in appendix \ref{app:coho}.

Thus, we have shown that the low dimensional characteristic classes of
$\clg$ bundles encode the Bianchi identities of massive IIA. That loop
group bundles seem to include the Romans mass (nonzero $G_0$)
indicates that there may be a larger story to be told here. We hope to
compute the full range of characteristic classes for $\clg$ bundles.
This is still under investigation.

\section{The cohomology of the classifying space}\label{sec:cohom}

In this section, we will show how the $\hlg$-bundle constructed above
fits into a larger structure of bundles and classifying spaces. We
then show that the various classes introduced in the previous section
are, in fact, characteristic classes of $\hlg$-bundles. We will also
show that $\Ht^*_{e}(X)$ computes the cohomology of the total space
of the circle bundle $Y$.

\subsection{A structure of classifying spaces}\label{sec:class}

First, we will prove the existence and commutativity of the diagram:
\begin{equation}
\label{bcd2}
\begin{split}
\begindc{\commdiag}[3]
        \obj(10,10)[bt1]{$\MB\BT$}
        \obj(30,10)[bhlg]{$\MB\hlg$}
        \obj(50,10)[x]{$X$}
        \obj(10,30)[t]{$\BT$}
        \obj(30,30)[bflg]{$\MB\flg$}
        \obj(50,30)[y]{$Y$}
        \obj(70,30)[bg]{$\MB E_8$}
        \mor{bhlg}{bt1}{$f$}
        \mor{x}{bhlg}{$g$}
        \mor{bflg}{bhlg}{}
        \mor{y}{x}{$\rho$}
        \mor{t}{bflg}{}
        \mor{y}{bflg}{$h$}
        \mor{y}{bg}{}
        \cmor((30,33)(50,37)(70,33))
             \pright(50,39){$i$} 
        \cmor((50,7)(30,3)(10,7))
              \pleft(30,1){} 
\enddc
\end{split}\ .
\end{equation}
In the process of doing so, we will encounter a number of other
interesting constructions that shed some light on the central
$\hlg$-bundle of this paper. In the above diagram, we begin with the
space $X$ and the circle bundle $Y$ over $X$. This bundle is
classified by the map from $X$ to $\MB\BT$. Over $Y$ there exists an
$E_8$-bundle classified by the map from $Y$ to $\MB E_8$.

We begin by constructing the map $h$. Let $p \in Y$ be a point and let
$\Sp = \rho^{-1}\rho(p) \subset Y$ be the fiber containing $p$. The
$E_8$-bundle, $E$ over $Y$, is trivial over $\Sp$, so the space of
sections $\Gamma(\Sp,E) \cong \mathrm{Maps}(\MS{1},E_8)$.  Assembling
these for all points $p$, we see that there is a natural basepoint,
given by $p$, and, thus, a global action by $\flg$. This $\flg$-bundle
over $Y$, which we will denote $E'$, defines the map $h$.

To construct $i$, we observe that $\flg$ fits into the following exact
sequence:
\begin{equation}
\label{blgexact}
1 \longrightarrow \blg \longrightarrow \flg \longrightarrow E_8 
\longrightarrow 1\ .
\end{equation}
The rightmost map in this sequence is the evaluation map
$\mathrm{ev}_p : \flg \rightarrow E_8$. We can use this to associate
an $E_8$-bundle to the $\flg$-bundle in the usual manner. We begin
with the product $E' \times E_8$. On this bundle, there is a
free $\flg$ action given by $(e,g) \rightarrow
(eh,\mathrm{ev}_p(h^{-1})g)$ for $h \in \flg$. The $\flg$ action on
$e$ is defined such that $p$, the projection of $e$ to the base, is
the base point allowing the $\flg$ action and also the point at which
the evaluation map acts.  The quotient by the $\flg$ action given an
$E_8$-bundle and defines the map $i$. On each fiber, the $\blg$
subgroup of $\flg$ acts only on the first factor in $E' \times E_8$
and reduces it to $E_p \times E_8$ where $E_p$ is the fiber of $E$ at
the point $p$. The quotient group $\flg/\blg \cong E_8$ acts freely on
$E_8$, so the quotient can be identified with $E_p$. This proves that
the composition of the maps $h$ and $i$ in (\ref{bcd2}) is the map
classifying the original $E_8$ bundle, $E$.

Now, let us examine the space $E'$, the total space of the
$\flg$-bundle over $Y$. This space has a natural $\flg$ action because
it is a principal bundle. There is also a natural $\BT$ action defined
by the simultaneous rotation of the circle in $\Gamma(\Sp,E)$ and the
circle in the fiber of $Y$ over $X$. By composing the two projections,
we can consider $E'$ as a fiber bundle over $X$ with fiber $\MS{1}
\times \flg$. The $\flg$ and $\BT$ actions just discussed combine to
give a free $\hlg$ action on $E'$. In other words, $E'$ is exactly the
$\hlg$-bundle given by the construction in section \ref{sec:bundle}.

Finally, $\hlg$ fits into the following 
exact sequence
\begin{equation}
\label{hlgexact}
1 \longrightarrow \flg \longrightarrow \hlg \longrightarrow \BT
\longrightarrow 1\ .
\end{equation}
In fact, this sequence splits on the right reflecting that $\hlg$ is a
semidirect product.  Now, let us consider the universal bundle,
$\ME\hlg$ over $\MB\hlg$. As $\flg$ is a subgroup of $\hlg$, it acts
on $\ME\hlg$ and we can take the quotient. As $\ME\hlg$ is
contractible and the action of $\flg$ is free, the quotient must be
$\MB\flg$. The remaining part of $\hlg$ is $\hlg/\flg \cong \BT$, so
$\MB\flg$ is a $\BT$-bundle over $\MB\hlg$. By the above, the pullback
of $\ME\hlg$ by $g$ is $E'$. Taking the quotient by $\flg$ gives the
space $\MB\flg$ and the map $h$ by construction. The vertical lines in
(\ref{bcd2}) are $\BT$ quotients.  This implies that $Y$, as a
circle bundle over $X$, is the pullback of the circle bundle $\MB\flg$
over $\MB\hlg$. Thus, $Y$ is classified by $fg$. This completes the
proof that the diagram commutes.

We can now demonstrate how to invert the construction of 
section \ref{sec:bundle}.
Given an $\hlg$-bundle $E'$ over $X$, we quotient by $\flg$
to obtain $Y$ and consider $E'$ as a $\flg$-bundle over $Y$. We can
then recover the original $E_8$-bundle over $Y$ as the associated
bundle to this $\flg$ bundle by the evaluation map as was done above.

\subsection{The ordinary classes}

Now, we turn to the characteristic classes. We begin by examining
those that exist in ordinary cohomology. As stated above, the origin
of the $E_8$ gauge field in M-theory is in the existence of a
four-form field strength $[G] \in H^4(Y,\BZ)$.  Because its homotopy
groups, $\pi_i(E_8)$, vanish for $i \neq 3$, $i \le 15$, for the
purposes of eleven dimensional manifolds, $E_8$ serves as a model for
$K(\BZ,3)$. Since the classifying space functor shifts homotopy groups
by one, this means that $\MB E_8$ is a model for $K(\BZ, 4)$. Finally,
given that $H^4(Y,\BZ)$ is equivalent to the homotopy class of maps
$[Y,K(\BZ,4)]$, we immediately see that $E_8$ bundles serve as models
for elements in $H^4(Y,\BZ)$ and, hence, for the field strength $G$.

The rational cohomology of these Eilenberg-MacLane spaces is well
known \cite{MR83i:57016}. This gives us the rational cohomology of the
classifying space $\MB E_8$ in low dimensions.  For dimensions less
than twelve, there is only a degree four element and its square. The
defining characteristic class of the $E_8$-bundle used in the previous
section is the pull back of the class in $H^4(\MB E_8,\BZ)$ which is,
of course, the same class as in the previous paragraph. Thus, this
class completely characterizes the $E_8$-bundle. One can also
construct this class via obstruction theory \cite{Witten:1986bt}
similarly to the class for $\clg$-bundles constructed in the previous
section. Finally, if we were to choose a connection on the
$E_8$-bundle, the four form can be given \`{a} la Chern-Weil as $\Tr(F
\wdg F)$.

We compute the rational cohomologies of $\MB\flg$ and $\MB\hlg$
in appendix \ref{app:coho}. The calculations involve the Leray-Serre
spectral sequence applied to various fibrations obtainable from the
exact sequences (\ref{blgexact},\ref{hlgexact}). The structure in
(\ref{bcd2}) also figures in prominently.

For $\MB\flg$, we obtain:
\begin{equation}
\label{bflgcoho}
\begin{array}{r||c|c|c|c|c|c|c|c|c|c|c|c}
i= & 0 & 1 & 2 & 3 & 4 & 5 & 6 & 7 & 8 & 9 & 10 & 11 \\
\hline
H^i(\MB\flg) & \BZ & 0 & 0 & x & y & 0 & 0 & xy & y^2 & 0  & 0 & xy^2 \\
\end{array}\ .
\end{equation}
Note that $x^2 = 0$ as it is of odd degree.

For $\MB\hlg$, we obtain.
\begin{equation}
\label{bhlgcoho}
\renewcommand{\arraystretch}{1.5}
\begin{array}{r||c|c|c|c|c|c|c|c|c|c|c}
i= & 0 & 1 & 2 & 3 & 4 & 5 & 6 & 7 & 8 & 9 & 10 \\
\hline
H^i(\MB\hlg) & \BZ & 0 & x & y & x^2 & 0 & x^3 & z & x^4 & 0 & x^6\\
\end{array}\ .
\end{equation}
In the cohomology ring we have the relations $xy =  xz = yz = 0$. 
Referring to (\ref{bcd2}), if $w$ is the generator of the cohomology
of $\MB\BT$, then $x = f^*(w)$.  

Also in (\ref{bcd2}), $g : X_{10} \rightarrow \MB\hlg$ is a classifying
map for the $\hlg$ bundle. We have the following identifications:
\begin{equation}
[G_2] = g^*(x) = (gf)^*(w) \qquad \mathrm{and}\qquad [H] = g^*(y)\ .
\end{equation}
The relation $xy = 0$ is reflected in the identity (\ref{g4ident}).

We also have the unfamiliar seven form $z$. Its proper 
interpretation, as we will demonstrate in the following subsection, is
\begin{equation}
\label{hg4rel}
2 [H \wdg G_4] = g^*(z)\ .
\end{equation}
This is a closed form, and it is not hard to see that it respects the
relations $xz = yz = 0$ in the cohomology ring. The factor of two may
seem mysterious, but it follows from the derivation (\ref{sqrtowdg}).
There are no other characteristic classes in dimensions less than
eleven.

The commutativity of the diagram (\ref{bcd2}) tells us that
the four form in M-theory is the pullback of the four form in
$\MB\flg$. The spectral sequence in the appendix tells us that the
pushforward of this form is the three form in $\MB\hlg$. This accords
with the fact that the three form in type IIA is the pushforward of
the M-theory four form.

Finally, we note that there is an interesting local construction of
$H$ which elucidates its interpretation in terms of obstruction
theory. To construct this class \cite{MR94b:57030}, choose a connection
$A$ on the $\hlg$ bundle $\pi: \Et \rightarrow X_{10}$. Let $\Lg$ and
$\cLg$ denote the Lie algebras of $\hlg$ and $\BT \ltimes \ulg$
respectively. Consider an open set $U$ in $X_{10}$ over which we can
locally lift the $\hlg$ bundle to a $\BT \ltimes \ulg$ bundle. Such a
lifting is a principal $\BT \ltimes \ulg$-bundle $p: \tilde{E}
\rightarrow U$ together with a $\hlg$ equivariant mapping $f:
\tilde{E} \rightarrow \Et$ such that $p = \pi(f)$. Choose a connection
$A_U$ on $\tilde{E}$ which is compatible with $A$ in the sense that
$f^*A = q(A_U)$, where $q: \cLg \rightarrow \Lg$ is the quotient map
that annihilates the generator of the central element $K$. Since,
$A_U \rightarrow A_U + \alpha K$ is still a compatible connection for
any one form $\alpha$ on $X_{10}$, the space of such connections is an
affine space under the space of one forms on $X_{10}$. Now, form the
curvature $F_U = d A_U + A_U \wedge A_U$ and note that $F_U
\rightarrow F_U + d\alpha K$. Define the scalar curvature $\Theta_U$
as the $K$ component of $F_U$. We define $H_U = d \Theta_U$. Clearly,
$H_U$ is independent of the choice of connection $A_U$ compatible with
$A$. Thus, it patches together globally over any open cover of
$X_{10}$ to a closed 3-form $H$. Further, if this form is
topologically nontrivial, the above local construction cannot be
extended to the whole manifold, and therefore it measures the 
obstruction to lifting the $\hlg$ bundle to a $\BT \ltimes \ulg$
bundle.

\subsection{The four form}

\begin{floatingfigure}
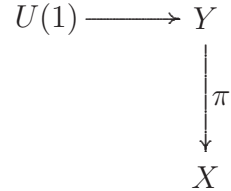
{2in}
\begindc{\commdiag}[30]
\obj(3,1)[B]{$X$}
\obj(1,3)[T]{$\BT$}
\obj(3,3)[E]{$Y$}
\mor{T}{E}{}
\mor{E}{B}{$\pi$}
\enddc
\caption{A circle bundle}
\label{circbundfig}
\end{floatingfigure}

Now, we would like to demonstrate a characteristic class in the
cohomology of the complex given in (\ref{doubform}). To do so, we will
show that the complex computes the cohomology of the total space, $Y$,
of a circle bundle over $X$ with Euler class, $e$, as in figure
\ref{circbundfig}.  Using the four form in $H^4(\MB\flg)$ from
(\ref{bflgcoho}), the fibration of $\MB\flg$ over $\MB\hlg$ in
(\ref{bcd2}) demonstrates the existence of the needed class. The
commutativity of (\ref{bcd2}) demonstrates that the class does encode
the correct $E_8$-bundle.

Rather than work with differential forms, we can define $\Ht^*_{e}$
over the integers by replacing the forms with singular cochains. As
such, for most of this section, we will consider cohomology with
integer coefficients. The goal is to prove:
\begin{equation}
\label{htiso}
        \Ht^*_{e}(X,\BZ) \cong H^*(Y,\BZ)
\end{equation}
We will prove this isomorphism by exploiting the Gysin sequence in
cohomology.  However, we would also like to show that the maps
introduced in the previous section, (\ref{g4def}) and (\ref{gdef}),
induce the above isomorphism when working with differential forms.
This can be accomplished by exhibiting a chain homotopy which is the
subject of appendix \ref{app:chain}.

Recall that the Gysin sequence is given by
\begin{equation}
\label{gysin}
\begindc{\commdiag}
\obj(1,1)[1]{}
\obj(3,1)[2]{$H^{i-2}(X)$}
\obj(5,1)[3]{$H^i(X)$}
\obj(7,1)[4]{$H^i(Y)$}
\obj(9,1)[5]{$H^{i-1}(X)$}
\obj(11,1)[6]{$H^{i+1}(X)$}
\obj(13,1)[7]{}
\mor{1}{2}{}
\mor{2}{3}{$\cup e$}
\mor{3}{4}{$\pi^*$}
\mor{4}{5}{$\pi_*$}
\mor{5}{6}{$\cup e$}
\mor{6}{7}{}
\enddc
\end{equation}
This implies that $H^i(Y)$ can be decomposed as follows:
\begin{equation}
\label{hisplit}
\begin{split}
H^i(Y) & \cong \pi^*\left[\mathrm{Coker}(\ \cup e) \subset H^i(X)\right] \\
            & \qquad \oplus \pi^{-1}_*\left[\mathrm{Ker}(\ \cup e) \subset H^{i-1}(X)\right]
\end{split}\ .
\end{equation}

To prove (\ref{htiso}), we will show that $\Ht_{e}^i(X)$ decomposes as
in (\ref{hisplit}). We begin by determining the space of all possible
inequivalent $\beta$ for $[(\alpha,\beta)] \in \Ht_{e}^i(X)$. We need
that $d\beta = 0$.  Furthermore, the invariance upon adding an exact
form $d(a,b)$ means that we have $\beta \sim \beta - db$. In other
words, the space of inequivalent $\beta$ is a subgroup of
$H^{i-1}(X)$. The condition that $d(\alpha,\beta)=0$ also implies that
for any cocycle representative $G_2$ of $e$,
$d\alpha + G_2\cup\beta = 0$. Therefore, we also require that
$e\cup[\beta] = 0$.  This tells us that the space of all possible $\beta$
is exactly $\mathrm{Ker}(\ \cup e) \subset H^{i-1}(X)$.

Once we fix a $\beta$, we have to ask which $\alpha$ exist such that
$[(\alpha, \beta)] \in \Ht_{e}^i(X)$. As we have $e \cup [\beta] = 0$,
we choose a $\gamma$ such that $d\gamma + G_2 \cup\beta = 0$. Any
element of $\Ht_{e}^i(X)$ can now be written as $(\alpha,0) +
(\gamma,\beta)$ with $d\alpha = 0$.  The only remaining exact forms
are those $d(a,b)$ with $db = 0$.  This preserves the subgroup of
pairs of the form $(\alpha,0)$. The invariance by adding $d(a,b)$
gives $\alpha \sim \alpha + da + e\cup b$ with $db = 0$.  This tells
us that the space of inequivalent $\alpha$ for a given $\beta \in
\mathrm{Ker}(\ \cup e) \subset H^{i-1}(X)$ is exactly
$\mathrm{Coker}(\ \cup e) \subset H^i(X)$ which demonstrates the
decomposition (\ref{hisplit}).

One might note now that the cohomology of $\MB\flg$ has a degree
three element and wonder how it appears in $\Ht^3_{e}(X_{10})$.
In fact, this three form, call it $H$, is just the pullback of 
the three form in $\MB\hlg$. Hence, it maps to the class
$(H,0)$ and is nothing new.

As a last exercise, we verify that the relation (\ref{hg4rel}) is
pulled back from the classifying space. For what follows, we will work
with differential forms.  Let $[(a,b)]$ be the element in
$\Ht^*_{e}(\MB\hlg)$ which pulls back to $[(G_4,H)]$. The form $c =
2(b \wdg a)$ is closed, and the addition of an exact `form' to $(a,b)$
adds an exact form to $c$, and, as such, does not affect its
cohomology class.  Finally, we need to demonstrate that $[c] = z$, the
seven form in $H(\MB\hlg)$. In fact, for a general principal circle
bundle with total space $Y$, base $X$ and projection $\pi$, this
construction gives a map $H^i(Y) \to H^{2i-1}(X)$. We claim that this
map is
\begin{equation}
\label{sqrmap}
[D] \mapsto  [\pi_*(D \wdg D)], \quad D \in \Omega^i(Y)\ .
\end{equation}
Then, the claim that $[c] = z$ follows from the spectral sequence
(\ref{bhlgspec}). There, the eight form in $H^*(\MB\flg)$ is the
square of the four form and is given by $\theta z$ in the $E_2$ term.
The push forward maps this $z$.

The addition of an exact form to $D$ does not affect the cohomology
class of $\rho_*(D \wdg D)$, so (\ref{sqrmap}) is an operation on
cohomology.  By the considerations of appendix \ref{app:chain}, we
know that, if $[D] \in H^*(Y)$ corresponds to $[(A,B)] \in
\Ht^*_{e}(X)$, then $[\rho^*(A) - \mcA \wdg \rho^*B] = [D]$. Squaring
this and pushing forward, we obtain
\begin{equation}
\label{sqrtowdg}
\begin{split}
\pi_*([D]^2) &= \left[ \pi_* \left(\pi^*A \wdg \pi^*A - 2 \mcA \wdg 
\pi^*B \wdg \pi^* A
        \right)\right] \\
                      &= -2 \left[ \pi_*\left(\mcA \wdg \pi^*B \wdg \pi^* 
A\right)\right] \\
                      &= -(-1)^B 2 \left[\left(B \wdg \pi_*(\mcA \wdg \pi^* 
A)\right)\right] \\
                      &= 2 \left[B \wdg A\right]
\end{split}
\end{equation}
giving the desired answer.
\newline
\section{Discussion}\label{sec:disc}

To conclude, we consider in greater detail the physical
implication of the quantization conditions suggested by the loop group
picture in IIA. We begin by considering the description of D8-branes
in this context, which requires an understanding of the relationship
between $\hlg$ and $\clg$ more clearly. We then consider briefly the
description of the other topological defects, \ie D4-, NS5-, and D6-branes, 
in the loop group picture. In particular, we find that the mysterious
torsional behavior of D6-brane charge in massive IIA has an
interesting physical interpretation in the context of Calabi-Yau
3-fold compactifications of IIA with $G_0 \neq 0$ as studied in
\cite{Polchinski:1996sm}. Furthermore, we consider an interesting
example in which the K-theory description of the quantization
condition at $g_s=0$ disagrees with our picture and comment on the
physical interpretation of the differences. 

Regions of massive IIA arise in string theory seperated by D8-brane
domain walls. In the loop group picture, a
D8-brane is a domain wall across which the fiber of the loop group
bundle changes \cite{JAU}.
Thus, it is interesting to understand the relation
between massive and massless IIA across such a domain wall. To
approach this question, it is useful to compare the loop groups $\hlg$
and $\clg$.  Topologically, $\hlg$ is $\MS{1} \times \flg$, which is
the trivial bundle over $\flg$. Thus, the various groups $\hlg$ and
$\clg$ exhaust all $\BT$-bundles over $\flg$. There is a problem with
this connection, however: the group structures of $\hlg$ and $\clg$
are very different. In particular, $\hlg$ fits into the exact sequence
\begin{equation}
1 \longrightarrow \flg \longrightarrow \hlg \longrightarrow \BT
\longrightarrow 1\ ,
\end{equation}
while $\clg$ fits into the sequence
\begin{equation} \label{centralex}
1 \longrightarrow \BT \longrightarrow \clg \longrightarrow \flg 
\longrightarrow 1\ .
\end{equation}
This reflects the fact that $\clg$ is a central extension of $\flg$
with $\BT$ as a normal subgroup while $\hlg$ is a semidirect product
where $\BT$ is \textit{not} a normal subgroup.

In fact, we can further form a semidirect product with $\BT$ for all
the groups $\clg$.  One might ask if the semidirect product circle is
necessary in this situation. Since bundles of these groups would all
have an additional 2-form characteristic class associated with the
extra $\BT$, they do not appear to be relevant.  Also, as we noted
earlier, the group $\hlg$ can be thought of as a subgroup of
$\mathrm{Diff}(\MS{1})\ltimes \flg$.  This is an example of the
Sugawara construction. In the centrally extended case, one
can form the semidirect product using the connected part of
$\mathrm{Diff}(\MS{1})$.
The significance of all of this is still obscure. It would be
interesting to understand if this picture of massive IIA is
connected to the relations of massive IIA to M-theory conjectured
by Hull \cite{Hull:1998vy} and further explored in \cite{Haack:2001iz,
  Moore:2002cp}.

Regardless, it is clear that in passing from massive IIA to massless
IIA, the geometric interpretation of the $p$-form field strengths is
radically altered. Locally, we expect that the forms $G_2$, $H$, and
$G_4$ are continuous across the domain wall. However, as their global
structures are quite different, it is not clear how their associated
topological charges, \ie  D6, NS5, and D4 brane charges, behave as we
cross a D8-brane.  Here, the geometric construction in terms of loop
group bundles can be of some use.\footnote{Some of this analysis
is also presented in \cite{JAU}.}  In particular, the
homotopy groups of the loop group fibers determine the possible
topological defects. Using the fact that topologically in low dimensions,
\begin{equation}
\hlg \cong \MS{1} \times \flg \cong \MS{1} \times E_8 \times \Omega
E_8 \approx K(\BZ,1) \times K(\BZ,2) \times K(\BZ,3)
\end{equation}
we see that the low dimensional homotopy groups of $\hlg$ are just,
\begin{equation}
\pi_1(\hlg) = \pi_2(\hlg) = \pi_3(\hlg) = \BZ.
\end{equation}
This means that $\hlg$ bundles admit topological defects carrying
$\BZ$ valued charges of codimension 3, 4, and 5, corresponding to the
D6, NS5, and D4 brane, as we would expect. Now, using the exact
homotopy sequence of the fibration of (\ref{centralex}), it is easy to
see that the low dimensional homotopy groups of $\clg$ are,
\begin{equation}
\pi_1(\clg) = \BZ_k \qquad \pi_2(\clg) = 0 \qquad \pi_3(\clg) = \BZ.
\end{equation}
Thus, we see that $\clg$ bundles admit a $\BZ_k$ charged, codimension
3 topological defect which we might call a massive D6-brane, no defect
corresponding to an NS5-brane, and a D4-brane. While it is natural to
expect that D4-brane charge is identified across a D8-brane domain
wall, the fate of the NS5-brane and D6-brane is less clear. However,
the obstruction theory interpretation of some of the characteristic
classes we have discussed can clarify matters somewhat. We saw that
$[H]$ in IIA can be identified as the obstruction to lifting the
corresponding $\hlg$ bundle to a $\BT \ltimes \ulg$ bundle. 
Since the NS5-brane sources an integral
$[H]$ flux through a linking 3-cycle, this suggests that the
associated $\hlg$ bundle cannot be lifted to a configuration in
massive IIA with $G_0=1$. Thus, we expect that the NS5-brane cannot
exist in massive IIA. This is, of course, clearly the right answer as
can be gleaned quite directly from the Bianchi identity of massive IIA,
$dG_2 = G_0 H$. 

The meaning of the $\BZ_k$ charge and the fate of D6-branes is
less obvious. In massive IIA, $[(G_2,H)]$ measures the obstruction to
lifting a $\clg$ bundle to a $\ulg$ bundle. Thus, this $\BZ_k$ charge
seems to be a defect that is nontrivial only if we have more than one
D8-brane domain wall present, suggesting an interpretation in terms of
the nonabelian gauge theory living on D8-branes. Since the defect is
of codimension 3, it also seems to be related to D6-branes. It
turns out \cite{Janssen:1999sa, Imamura:2001cr} that there exists a supergravity solution of
massive IIA called the massive D6-brane which has been interpreted as
a D8-D6-NS5 intersection, with the NS5-brane stuck to the domain wall.
However, it is not clear why these defects carry a discrete torsional
charge from this perspective. Furthermore, the precise connection to
D6-branes in massless IIA is still not obvious.\footnote{Other
interesting bound states have been explored in 
\cite{Singh:2001gt,Singh:2002eu}.}

While we will not be able to completely resolve this issue here, we
will motivate an answer based on an interesting physical
interpretation of the results of the previous paragraphs. It turns out
that one can understand the behavior of NS5-branes and
D6-branes in massive IIA via a kind of St\"{u}ckelberg mechanism. In
massive IIA, the gauge transformation of the NSNS $B$-field acts on the
RR gauge potential $C_1$ as well,
\begin{equation}
  \delta C_1 = -G_0 \Lambda_{NS} \qquad \delta B = d \Lambda_{NS}.
\end{equation}
In particular, this suggests that we can locally gauge away the $C_1$
field in massive IIA. The interpretation (see, for example, \cite{Janssen:1999sa}) for
this is that the $B$-field actually becomes massive by eating $C_1$.
In other words, $C_1$ is a St\"{u}ckelberg field, and its kinetic energy term
becomes the mass term for $B$ in unitary gauge. Under
electric/magnetic duality, the dual RR $C_7$ field becomes massive by
eating the dual NSNS $B_6$ field. Thus, we might expect that at $g_s
\neq 0$, regions of massive IIA repel magnetic $B$-fields, and, in
particular, free magnetic monopoles such as NS5-branes would 
not be present. Furthermore, as the dual RR $C_7$ field is massive, one
might expect that the vacuum of the theory has a condensate carrying
its electric charge. In particular, this is the charge carried by a
D6-brane, and if the condensate is a coherent state constructed
out of quanta with charge $k$, we might expect that the D6-brane
charge is only conserved modulo $k$.

The work of \cite{Polchinski:1996sm} provides a nearly ideal
example for testing the above interpretation. They consider Calabi-Yau
compactifications of massive IIA down to four space-time
dimensions. There, the axio-dilaton acquires a nonzero RR 1-form
magnetic charge of precisely $G_0$ units.\footnote{This connection was
actually used to justify the quantization condition on $G_0$.} Thus,
if we are at $g_s \neq 0$, leading the axio-dilaton field to condense,
we see that this charge, which is precisely a D6-brane charge, is only
conserved modulo $G_0$. In fact, they show that near the conifold
point, this behavior can be attributed to the condensation of
D6-branes wrapped on the CY 3-fold.  Note, however, that at precisely
$g_s=0$, this interpretation fails, and D6-brane charge should be
$\BZ$ valued. We believe that this lends support to the notion that
the $\clg$ quantization condition should be interpreted as an
important ingredient in describing a kind of M-theoreric, finite $g_s$
dual to massive IIA.

Finally, we would like to make a more direct comparison between the loop
group picture and K-theory on certain simple manifolds. Consider type
IIA compactified to 6-dimensions on $X=\CP{1} \times \CP{1}$ (if we
wish to, we can further compactify the remaining six dimensions on
a six-manifold with positive scalar curvature, for example, a
six-sphere). We will be interested in analyzing nontrivial $G_2$ and
$G_4$ flux configurations on $X$ corresponding to the K-theory classes
of the the virtual bundles, $x_n=\mcO(n,n) \ominus \mcO(n-1,n-1)$. As
$X$ and all its factors have trivial normal bundle and positive scalar
curvature, the arguments of section 4 of \cite{Moore:1999gb} show that
there is no nontrivial shift in the quantization conditions. Furthermore,
as these virtual bundles have rank $0$, if $x$ and $y$ are the
generators of the cohomology of the two $\CP{1}$ fibers, we have,
\begin{equation}
\begin{split}
[G(x_n)] = [G_2] + [G_4]  & = \sqrt{\hat{A(X)}} ch(x_n) = ch(x_n) \\
& = e^{n(x+y)} - e^{(n-1)(x+y)} = (x+y) + (n- \frac{1}{2})(x+y)^2\ .
\end{split}
\end{equation}
As their Chern characters are clearly different, the classes $x_n$ are
distinguishable and correspond to distinct allowed flux configurations
on $X$. What is interesting in this example is that, as $n$ varies, the
K-theory classes have the same $[G_2]$ while 
their values of $[G_4]$ differ
by integral multiples of $[\alpha \wedge G_2]$ for some closed
$\alpha$ (here, $\alpha = G_2$). In particular, their differences are
trivial as elements of $\Ht_e^4(X)$ and lift to the same
class $a$ in M-theory. This is, then, a simple example in which the
K-theory description of the quantization of fluxes seems to be at odds
with that of M-theory even in the absence of torsion. Note, however,
that while the topological data of the $E_8$-bundle in M-theory is
insensitive to the difference between these configurations, the same
is not true for the CS terms in the M-theory action. Thus, this data
is certainly detected by the $\eta$ invariant of the $E_8$-bundle.
While we will not be able to resolve this puzzle, we can see
two possible resolutions. As we saw in the previous example, it is
possible that finite $g_s$ effects, perhaps non-perturbative instanton
effects, may be responsible for a breakdown in the K-theory
picture. However, the authors of \cite{Diaconescu:2000wy} were able to escape
this difficulty by considering their $E_8$-bundles as pulled back from
the ten dimensional base, giving a finer characterization.
This suggests that there may be some equivariant information that
could distinguish $E_8$-bundles.

Let us end by briefly touching on the issue of twisted K-theory.
We can combine the form fields present in supergravity
with their Hodge duals to generate the self-dual form:
\begin{equation}
G_t = G_0 + G_2 + G_4 + \star G_4 + \star G_2 + \star G_0 \ .
\end{equation}
The Bianchi identities and the equations of motion imply that
\begin{equation}
dG_t + H\wdg G_t = 0\ .
\end{equation}
If we define an operation on an arbitrary form, $F$, by
\begin{equation}
d_H(F) = dF + H \wdg F\ ,
\end{equation}
we have $d_H^2 = 0$ and can take its cohomology. This is sometimes
termed \textit{twisted cohomology} and is denoted $H(X_{10},[H])$ 
\cite{Rohm:1986jv}.  This cohomology is tantalizingly similar, but in many
ways opposite, to the generalized cohomology defined in the text.

It is believed that RR forms in IIA, rather than being elements in
this cohomology, however, actually live in the twisted K-theory group
$K^0_{[H]}(X_{10})$ \cite{Bouwknegt:2000qt,Bouwknegt:2001vu}.  There
is a Chern character that maps this group to the twisted cohomology
defined above \cite{Mathai:2002yk}. The relation between twisted
cohomology, twisted K-theory and $E_8$ gauge theory has been explored
in \cite{Mathai:2003mu,Evslin:2003hd}.


If it turns out that $\hlg$ bundles fully capture the information of 
the 4-form in M-theory, we would have two alternate 
quantizations of the fields of type IIA supergravity. One might hope 
that this could shed some light on the calculations of 
\cite{Diaconescu:2000wy}.

\section{Acknowledgments}

We are indebted to Dan Freed for his assistance throughout this paper.
We would also like to thank David Ben-Zvi and Jacques Distler for
useful conversations. UV would like to thank Allan Adams and Jarah
Evslin for stimulating conversations, and Petr Ho\v{r}ava for
suggesting this line of inquiry to us.  This material is based upon
work supported by the National Science Foundation under Grant No.
PHY-0071512.

\appendix\section{Some cohomology computations}\label{app:coho}

\subsection{$\MB\flg$}
First, we compute the cohomology of $\MB\flg$. We recall the 
exact sequence
\begin{equation}
\label{blgexact2}
1 \longrightarrow \blg \longrightarrow \flg \longrightarrow E_8 
\longrightarrow 1
\end{equation}
where the last map is given by evaluation at the basepoint of $\MS{1}$. 
Similarly to the construction of the bundle circle bundle
over $\MB\hlg$ in section \ref{sec:class},
we can form the principal bundle:
\begin{equation}
\label{classfiber}
\begin{split}
\begindc{\commdiag}
        \obj(3,1)[bflg]{$\MB\flg$}
        \obj(3,3)[bblg]{$\MB\blg$}
        \obj(1,3)[g]{$E_8$}
        \mor{g}{bblg}{}
        \mor{bblg}{bflg}{}
\enddc\ .
\end{split}
\end{equation}
Now, we can look at the Leray-Serre spectral sequence\footnote{See,
  for example, \cite{MR83i:57016,MR2002c:55027}.} for
(\ref{classfiber}).  The Whitehead theorem tells us that the
cohomology of both $\MB\blg$ and $E_8$ is that of $K(\BZ,3)$ in low
dimensions.  If we denote the generator of $H^3(K(\BZ,3),\BZ)$ by
$\theta$, the only choice we have is whether or not $d_4 \theta = 0$.
As the Hurewicz isomorphism tells us that $H^3(\MB\flg,\BZ) = \BZ$, we
must kill $\theta$, implying the existence of an element $y$ such that
$d_4 \theta = y$. Thus, we can write down the $E_2\ (=E_3 = E_4)$ term
as
\begin{equation}
\label{bflgspec}
\begin{split}
\begin{array}{r||c|c|c|c|c|c|c|c|c|c|c|c}
3 & \theta &  & & \theta x & \theta y & & & \theta xy & \theta y^2 & & 
& \theta xy^2 \\
\hline
2 & & & & & & & & & & & & \\
\hline
1 & & & & & & & & & & & & \\
\hline
0 & \BZ & & & x & y & & & xy & y^2 & & & xy^2 \\
\hline
\hline
& 0 & 1 & 2 & 3 & 4 & 5 & 6 & 7 & 8 & 9 & 10 & 11 \\
\end{array}
\end{split}\ .
\end{equation}
After we take the cohomology of $d_4$ to obtain $E_5$, we see that the
spectral sequence collapses, at least for the low dimensions that we
care about.  The only terms that survive are $\BZ$ and $x$ giving the
rational cohomology of $K(\BZ,3)$. The zero row of (\ref{bflgspec}) is
the cohomology of $\MB\flg$ yielding the result (\ref{bflgcoho}). In
fact, one can see directly that $\MB\flg$ is approximately $K(\BZ,3)
\times K(\BZ,4)$ in low dimensions, but we will not do so here.

\subsection{$\MB\hlg$}

We now compute the cohomology of $\MB\hlg$. We begin by recalling the
diagram \ref{bcd2}:
\begin{equation}
\label{bigcommdiag}
\begin{split}
\begindc{\commdiag}[3]
        \obj(10,10)[bt1]{$\MB\BT$}
        \obj(30,10)[bhlg]{$\MB\hlg$}
        \obj(50,10)[x]{$X$}
        \obj(10,30)[t]{$\BT$}
        \obj(30,30)[bflg]{$\MB\flg$}
        \obj(50,30)[y]{$Y$}
        \obj(70,30)[bg]{$\MB E_8$}
        \mor{bhlg}{bt1}{}
        \mor{x}{bhlg}{}
        \mor{bflg}{bhlg}{}
        \mor{y}{x}{}
        \mor{t}{bflg}{}
        \mor{y}{bflg}{}
        \mor{y}{bg}{}
        \cmor((30,33)(50,38)(70,33))
             \pright(50,40){$i$}
\enddc
\end{split}\ .
\end{equation}

First, note that there exists a two form pulled back from $\MB\BT$. As
any circle bundle can exist as a part of the $\hlg$-bundle, any map to
$\MB\BT$ factors through $\MB\hlg$.  This reflects that the sequence
(\ref{hlgexact}) splits. This universality implies that no power of
the two form in $\MB\hlg$ can vanish.

Next, we isolate the following fibration from (\ref{bigcommdiag}):
\begin{equation}
\begin{split}
\begindc{\commdiag}
         \obj(3,1)[Bhlg]{$\MB\hlg$}
         \obj(1,3)[T]{$\BT$}
         \obj(3,3)[Bflg]{$\MB\flg$}
         \mor{Bflg}{Bhlg}{}
         \mor{T}{Bflg}{}
\enddc
\end{split}\ .
\end{equation}
When constructing the $E_2$ term of the associated spectral
sequence, there are two possibilities. The question is whether
the four form of $\MB\flg$ is pulled back from $\MB\hlg$ or not.
We will see that it is not. Thus, the $E_2$ term is
\begin{equation}
\label{bhlgspec}
\begin{split}
\begin{array}{r||c|c|c|c|c|c|c|c|c|c|c}
1 & \theta & & \theta x & \theta y & \theta x^2 & & \theta x^3 & \theta 
z & \theta x^4
&  & \theta x^6 \\
\hline
0 & \BZ & & x& y & x^2 & & x^3 & z & x^4 &  & x^6\\
\hline
\hline
& 0 & 1 & 2 & 3 & 4 & 5 & 6 & 7 & 8 & 9 & 10\\
\end{array}
\end{split}
\end{equation}
where all products of $x$, $y$ and $z$ are zero. As above, the
zero row gives the cohomology (\ref{bhlgcoho}). Note that
the four form in $\MB\flg$ is $\theta x$ which pushes forward
to $x$, the three form in $\MB\hlg$.

In order to prove the result about the four form, 
we proceed by examining the following fibration
which follows from (\ref{blgexact},\ref{blgexact2}):
\begin{equation}
\begin{split}
\begindc{\commdiag}
        \obj(3,1)[bg]{$\MB E_8$}
        \obj(3,3)[bflg]{$\MB\flg$}
        \obj(1,3)[bblg]{$\MB\blg$}
        \mor{bflg}{bg}{$i$}
        \mor{bblg}{bflg}{}
\enddc
\end{split}\ .
\end{equation}
The $E_2$ term in the associated spectral sequence is as follows:
\begin{equation}
\begin{split}
\begin{array}{r||c|c|c|c|c}
3 & \theta & & & & \theta x \\
\hline
2 & & & & & \\
\hline
1 & & & & & \\
\hline
0 & \BZ & & & & x \\
\hline
\hline
& 0 & 1 & 2 & 3 & 4 \\
\end{array}
\end{split}\ .
\end{equation}
As we know the total cohomology already from the previous
section, we see that there are no nontrivial differentials.
This implies that the four form in $\MB\flg$ is pulled back
from the four form in $\MB E_8$ by $i$.

Any four form in $Y$ can be represented by a map to $\MB E_8$. The
commutativity of (\ref{bigcommdiag}) implies that this four form can
be thought of as pulled back from $\MB\flg$.  Now, we assume, as
above, that the four form in $\MB\flg$ is \textit{also} pulled back
from $\MB\hlg$.  By commutativity, we could then pull it back to $Y$
through $X$, giving the same form as when pulled back from $\MB E_8$.
Now, let $Y \cong \MS{3} \times \MS{1}$ and $X \cong \MS{3}$.  There
exists a nontrivial four form on $Y$ which clearly cannot be pulled
back from $X$. Thus, we have a contradiction, so the four form on
$\MB\flg$ is not pulled back from $\MB\hlg$.

\subsection{$\MB\clg$}

As a last exercise, we prove the existence of a four form
characteristic class in $\MB\clg$. First, we recall that
$\clg$ fits into the following fibration:
\begin{equation}
\begin{split}
\begindc{\commdiag}
        \obj(3,1)[flg]{$\flg$}
        \obj(3,3)[clg]{$\clg$}
        \obj(1,3)[T]{$\BT$}
        \mor{T}{clg}{}
        \mor{clg}{flg}{}
\enddc
\end{split}\ .
\end{equation}
In low dimensions, the homotopy sequence of the fibration
gives the following set of homotopy groups for $\clg$:
\begin{equation}
\begin{array}{r||c|c|c|c|c|c|c}
i= & 0 & 1 & 2 & 3 & 4 & 5 & 6 \\
\hline
\pi_i(\clg) & \BZ & \BZ_p & 0 & \BZ & 0 & 0 & 0 \\
\end{array}\ .
\end{equation}
In fact \cite{MR88i:22049}, for $k=1$, the group is simply connected,
and for higher $k$, $p=k$.  The higher extensions are all $\BZ_k$
quotients of the level one extension. Finally, it follows from the
rational Hurewicz isomorphism that $H^4(\MB\clg,\BQ) = \BQ$.

\section{The chain homotopy}\label{app:chain}

In this appendix, we will show that the equations (\ref{g4def}) and
(\ref{gdef}) induce an isomorphism between $\Ht^*_{e}(X) \cong
H^*(Y)$. First, we examine the following diagram:
\begin{equation}
\begin{split}
\begindc{\commdiag}[3]
\obj(10,10)[s2]{}
\obj(10,30)[s1]{}
\obj(30,10)[o12]{$\Ot^i(X)$}
\obj(30,30)[o11]{$\Omega^i(Y)$}
\obj(50,10)[o22]{$\Ot^{i+1}(X)$}
\obj(50,30)[o21]{$\Omega^{i+1}(Y)$}
\obj(70,10)[e2]{}
\obj(70,30)[e1]{}
\mor{s1}{o11}{}
\mor{o11}{o21}{$d$}
\mor{o21}{e1}{}
\mor{s2}{o12}{}
\mor{o12}{o22}{$d$}
\mor{o22}{e2}{}
\mor(31,30)(31,10){$j$}
\mor(51,30)(51,10){$j$}
\mor(29,10)(29,30){$j^{-1}$}
\mor(49,10)(49,30){$j^{-1}$}
\enddc
\end{split}
\end{equation}
where $j(\alpha) = (-1)^\alpha(\rho_*(\mcA \wdg \alpha),\rho_*(\alpha
))$ and $j^{-1}(\beta,\gamma) = \rho^*(\beta) - \mcA \wdg
\rho^*(\gamma)$, essentially the maps (\ref{g4def}) and (\ref{gdef}).
In $(-1)^\alpha$, $\alpha$ is just the degree of the form $\alpha$ and
the multiplication is distributive across the direct sum.  It is easy
to verify that $j $ and $j^{-1}$ are chain maps. As always, we choose
a specific representative of the Euler class, $e$, and a connection,
$\mcA$, that satisfy the relations (\ref{connecteqn}).


Next, we must show that the induced maps on cohomology are actually
isomorphisms. First, we show that $j\circ j^{-1}$ is the identity on 
the level of forms:
\begin{equation}
\begin{split}
(j\circ j^{-1})(\beta,\gamma) &= j(\rho^*(\beta) -
\mcA\wdg\rho^*(\gamma))\\
          &= (-1)^\beta (\rho_*(\mcA \wdg (\rho^*(\beta) - \mcA\wdg
          \mcA\wdg\rho^*(\gamma))),
                 \rho_*(\rho^*(\beta) - \mcA\wdg\rho^*(\gamma)) )\\
           &= (\rho_*(\mcA) \cdot \beta, \rho_*\rho^*(\beta) + 
\rho_*(\mcA)
           \cdot \gamma) \\
           & = (\beta, \gamma)\ .
\end{split}
\end{equation}

Now, let us examine the situation for $j^{-1}\circ j$:
\begin{equation}
\begin{split}
\label{toughinverse}
(j^{-1}\circ j)(\alpha) &= (-1)^\alpha
j^{-1}(\rho_*(\mcA\wdg\alpha),\rho_*(\alpha)) \\
      &= (-1)^\alpha \left[\rho^*\rho_*(\mcA\wdg\alpha) -
      \mcA\wdg\rho^*\rho_*(\alpha)\right]\ .
\end{split}
\end{equation}
This is clearly not equal to $\alpha$. In order to go further, we must
investigate the properties of the operator $\rho^*\rho_*$. Let us
write $\alpha$ in local coordinates as
\begin{equation}
\label{alphaloc}
\alpha = f^i(\theta,\mathbf{x}) d\mathbf{y}_i \wdg d\theta +
g^i(\theta,\mathbf{x}) d\mathbf{z}_i
\end{equation}
where $d\mathbf{y}_i$ and $d\mathbf{z}_i$ are bases of forms of the
appropriate degree on the base pulled back by the fiber projection.
The sums over $i$ are understood. Then, we have:
\begin{equation}
\rho^*\rho_*(\alpha) = \left(\int d\theta f^i(\theta,\mathbf{x})\right)
d\mathbf{y}_i\ .
\end{equation}
We can also write $\mcA$ in local coordinates as:
\begin{equation}
\mcA = \frac{1}{2\pi} d\theta + \rho^*(\psi)\ .
\end{equation}
This follows because the connection is invariant with respect to the
$\BT$ action on the fiber implying that there cannot be any $\theta$
dependence in the coefficient of $d\theta$ or in the basic form
$\rho^*(\psi)$.  Furthermore the normalization condition in
(\ref{connecteqn}) forces the coefficient of $d\theta$ to be $1/2\pi$.
We also note that (\ref{toughinverse}) is invariant under additions to
$\mcA$ of the form $\rho^*(\psi)$ so we can neglect that term.
Finally, (\ref{toughinverse}) is linear on $\alpha$ and vanishes for
$\alpha = \rho^*(\beta)$.

We can put all this together to obtain:
\begin{equation}
(j^{-1}\circ j)(\alpha) = \left(\frac{1}{2\pi}\int d\theta g^i\right)
d\mathbf{z}_i + \left(\frac{1}{2\pi}\int d\theta f^i\right) 
d\mathbf{y}_i
\wdg d\theta
\end{equation}
Again, from the linearity of (\ref{toughinverse}), we can deal with
the terms in (\ref{alphaloc}) separately. Let us define an operator,
$K$, as follows:
\begin{equation}
\begin{split}
K( f^i(\theta,\mathbf{x}) d\mathbf{y}_i \wdg d\theta) & = \tilde{f^i}
d\mathbf{y}_i\\
K(g^i(\theta,\mathbf{x}) d\mathbf{z}_i) &= 0
\end{split}\ .
\end{equation}
Here $\tilde{f^i}$ is the function defined such that:
\begin{equation}
\pxpy{\tilde{f}^i}{\theta} = f^i - \frac{1}{2\pi}\int d\theta f^i
\end{equation}
and
\begin{equation}
\int d\theta \tilde{f}^i = 0
\end{equation}
which can always be achieved by adjusting the constant of integration.

Finally, we look at $(dK - Kd)\alpha$:
\begin{equation}
\begin{split}
(dK - Kd)\alpha &= d(\tilde{f} d\mathbf{y}_i)
         - K\Bigg[\left(\pxpy{\tilde{f}^i}{\mathbf{x}_j}d\mathbf{x}_j 
\right)
         \wdg d\mathbf{y}_i
         \wdg d\theta + (-1)^\alpha \pxpy{g^i}{\theta} d\mathbf{z}_i \wdg
         d\theta \\
         & \qquad\qquad  +
         \left(\pxpy{g^i}{\mathbf{x}_j}d\mathbf{x}_j\right)\wdg 
d\mathbf{z}_i\Bigg] \\
     &= \left[\pxpy{\tilde{f}}{\mathbf{x}_j}d\mathbf{x}_j\wdg 
d\mathbf{y}_i
           + (-1)^{\alpha - 1} \left(f^i - \frac{1}{2\pi}\int d\theta
           f^i\right) d\mathbf{y}_i \wdg d\theta\right] \\
     & \qquad\qquad - \widetilde{\pxpy{f^i}{\mathbf{x}_j}}
         d\mathbf{x}_j \wdg d\mathbf{y}_i - (-1)^\alpha
         \widetilde{\pxpy{g^i}{\theta}} d\mathbf{z}_i\\
     &= (-1)^{\alpha-1}\left[\left(f^i - \frac{1}{2\pi}\int d\theta f^i
     \right) d\mathbf{y}_i d\theta
         + \left(g_i - \frac{1}{2\pi}\int d\theta g_i\right)
         d\mathbf{z}_i\right] \\
     &= (-1)^{\alpha-1} (1 - (j^{-1} \circ j)(\alpha))\ .
\end{split}
\end{equation}
where we have used the fact that the tilde operation commutes with
differentiating on variables that are not $\theta$ and the following:
\begin{equation}
\pxpy{}{\theta}\widetilde{\pxpy{g^i}{\theta}} = \pxpy{g^i}{\theta}
\Longrightarrow
\widetilde{\pxpy{g^i}{\theta}} = g^i - \int d\theta g^i\ .
\end{equation}
This shows that $K$ is a homotopy operator, and therefore
$j$ and $j^{-1}$ do induce isomorphisms in cohomology.

\bibliographystyle{utphys}
\bibliography{thebib}
\end{document}